\documentclass[printer]{aa}  

\usepackage{graphicx}
\usepackage{amsmath} 
\usepackage{amssymb}
\usepackage{lscape}
\usepackage{longtable,lscape}
\usepackage{aalongtable,lscape}
\usepackage[sort&compress]{natbib}
\usepackage{natbib}
\usepackage{pifont}

\bibpunct{(}{)}{;}{a}{}{,}

\usepackage{txfonts}

\begin{document}

\title{Searching for planetary-mass T-dwarfs in the core of Serpens}

\subtitle{   }

\author{L. Spezzi\inst{1} \and C. Alves de Oliveira\inst{2} \and E. Moraux\inst{3}  \and J. Bouvier\inst{3} \and  E. Winston\inst{4} \and  P. Hudelot\inst{5} \and H. Bouy\inst{6} \and  J.-C. Cuillandre\inst{7}}

 \offprints{L. Spezzi, \email{lspezzi@eso.org}}
 
\institute{European Southern Observatory, Karl-Schwarzschild-Strasse 2, 85748 Garching bei M\"unchen, Germany
\and Herschel Science Centre, European Space Astronomy Centre (ESA), P.O. Box, 78, 28691 Villanueva de la Ca\~{n}ada, Madrid, Spain
\and UJF-Grenoble 1 / CNRS-INSU, Institut de Plan\'etologie et d'Astrophysique de Grenoble (IPAG) UMR 5274, 38041 Grenoble, France  
\and European Space Agency (ESTEC), PO Box 299, 2200 AG Noordwijk, the Netherlands
\and Institut d'Astrophysique de Paris, UMR 7095 CNRS, Universit\'e Pierre et Marie Curie, 98bis boulevard Arago, 75014 Paris, France 
\and Centro de Astrobiolog\'ia (INTA-CSIC), LAEFF, PO Box 78, 28691 Villanueva de la Ca\~nada, Spain 
\and Canada-France-Hawaii Telescope, 65-1238 Mamalahoa Hwy, Kamuela, HI 96743}

\date{Received 9 May 2012; accepted 1 August 2012}

 
  \abstract
   {The knowledge of the present-day mass function of young clusters and the mass of their coolest substellar members is essential to clarify 
   the brown dwarf formation mechanism, which still remains a matter of debate. }  
   {We searched for isolated planetary-mass T-dwarfs in the $\sim$3~Myr old Serpens Core cluster. }
   {We performed a deep imaging survey of the central part of this cluster using the WIRCam camera at the CFHT. Observations were performed through the 
   narrow-band CH$_4$off and CH$_4$on filters, to identify young T-dwarfs from their 1.6$\mu$m methane absorption bands, and the broad-band $JHK_S$ filters, 
   to better characterize the selected candidates. We complemented our WIRCam photometry with optical imaging data from MegaCam at CFHT and Suprime-Cam at the Subaru telescope and mid-infrared flux measurements from the Spitzer  ``core to disk'' (c2d) Legacy Survey.}
   {We report four faint T-dwarf candidates in the direction of the Serpens Core with CH$_4$on$-$CH$_4$off above 0.2~mag, estimated visual extinction 
   in the range 1-9~mag and spectral type in the range T1-T5 based on their dereddened CH$_4$on$-$CH$_4$off colors. 
   Comparisons with T-dwarf spectral models and optical to mid-infrared color-color and color-magnitude diagrams, 
   indicate that two of our candidates (ID~1 and 2) are background contaminants (most likely heavily reddened low-redshift quasars). 
   The properties of the other two candidates (ID~3 and 4) are consistent with them being young members of the Serpens Core cluster, 
   although our analysis can not be considered conclusive. In particular, ID~3 may also be a foreground T-dwarf. 
   It is detected by the Spitzer c2d survey but only flux upper limits are available above 5.8~$\mu$m and, hence, 
   we can not assess the presence of a possible disk around this object. However,  it presents some similarities with other young 
   T-dwarf candidates (S~Ori~70 in the $\sigma$~Orionis cluster and CFHT\_J0344+3206 in the direction of IC~348). 
   If ID~3 and 4 belong to Serpens, they  would have a mass of a few Jupiter masses and would be amongst the youngest, 
   lowest mass objects detected in a star-forming region so far.}


\maketitle

\section{Introduction}

Understanding the brown dwarf (BD) formation mechanism is a key point to assess what determines and which is the minimum mass for star formation.
Although two decades have passed since the first unambiguous observations of BDs \citep{Reb95,Nak95}, 
this issue is still under debate. According to the star formation theory, the minimum mass for star formation is set by the so called mass fragmentation limit, 
i.e. the mass at which one object cannot contract further because the radiated heat prevents it to collapse further. 
This limit is of the order of a 5-10M$_J$ \citep{Low76,Ree76} and may be lower ($\sim$1~M$_J$) in the presence of magnetic fields \citep{Bos01,Boy05}.
Observationally, several young star forming regions have been probed to faint magnitudes at both optical and near-infrared wavelengths 
to look for a possible cut-off at the low-mass end of the mass function and the results significantly vary from region to region. 
A possible cut-off below $\sim$6~M$_J$ up to 20~M$_J$ have been reported in NGC~1333 \citep[e.g.,][]{Sch09,Sch12}. 
Moreover, there is no universal agreement on the behavior of the mass function over this cut-off. A search for T-dwarfs in IC~348 by \citet{Bur09} suggests 
that an extrapolation of the log-normal mass function \citep{Cha03} may hold in the substellar and planetary-mass regimes. 
A similar study in the Upper Sco association favors a turn-down in the mass function below 10~M$_J$ \citep{Lod11}. 
In the Pleiades \citet{Cas07} reported several L/T-type candidates 
with masses as low as 10~M$_J$, implying that the slope of the mass function in this mass regime agrees, within the uncertainties, 
with the values inferred from earlier studies at higher masses \citep{Dob02,Mor03,Lod07,Cas07}.
Finally, \citet{Kir12} find tantalizing hints that the number of BDs in the field continues to rise from late-T to early-Y.

To address these questions, a large observing key-program (P.I. J. Bouvier) was conducted at the Canada France Hawaii Telescope (CFHT) 
aimed at characterizing the stellar population of nearby star forming regions using the 
Wide Field IR Camera (WIRCam). One of the main objectives of this program is to look for the lowest mass 
members of these clusters and investigate the cut-off at the low mass end of the mass function,  
using a technique based on narrow-band methane imaging \citep[e.g.,][]{Bur09}. 

In this paper, we report the results of this observing program in the Serpens Core cluster, 
specifically focusing on the methane filters imaging to search for young T-dwarfs candidates. 
The Serpens Core region is an example of a very young, deeply embedded (A$_V \approx$40~mag) cluster, containing a high percentage of protostars 
\citep{Dav99,Tes00,Kaa04,Eir06,Har07,Win07}. With an age between 2 and $\sim$6~Myr \citep{Win09,Oli09} 
and a distance estimated between 260$\pm$37~pc \citep{Str03} and 415$\pm$5~pc \citep{Dzi10}, 
the Serpens cloud core is one of the nearest regions of clustered 
star formation to the Sun. Therefore, it is an excellent candidate for study as it is close enough to both resolve the individual 
members and detect the lowest mass members to below the hydrogen-burning limit. 
Throughout this paper, we assume a typical age of 3~Myr \citep{Win09} for the Serpens core cluster and the minimum distance of 260$\pm$37~pc 
 \citep{Str03} \footnote{This choice is appropriate because we use the distance to Serpens 
 to estimate the lower limit to the interstellar extinction towards the cloud (see also Sect.~\ref{sect_par}).}. 
The region is also particularly interesting for BD studies because only a few dedicated searches have been reported in the literature. 
The first young BD in Serpens was discovered by \citet{Lod02} and the cluster substellar population count so far
about 3 spectroscopically confirmed BDs  \citep{Lod02,Shi11} and 40 BD candidates, 
a considerable fraction of which still surrounded by prominent accretion disks \citep{Eir06}.

Canonical approaches on BD studies and related issues \citep[see, e.g.,][]{Jay03} rely on the identification and spectral classification of 
BDs. However, optical/near-infrared (IR) spectroscopy \citep[see, e.g.,][]{McL03,Geb02} 
has proved impossible for very faint distant BDs with current telescopes and very time-consuming for large BD samples.
To overcome these limitations, new BD classification methods have been devised on the basis of 
near-IR broad-band \citep[see, e.g.,][and references therein]{Leg10,Sch10} and narrow-band imaging 
that targets unique molecular features of L and T dwarfs, in particular, water and methane bands \citep[see, e.g.,][]{Gor03,Mai04,Spe11}. 
These techniques are equivalent to extremely low-resolution spectroscopy and can be confidently applied 
for statistical purposes, e.g., to detect and classify \emph{bona fide} BDs in large imaging surveys.
Our search for T-dwarfs in Serpens is based on narrow-band methane (CH$_4$) imaging. We target, in particular, 
the CH$_4$ absorption band centered at 1.66~$\mu$m, which forms in the atmosphere of objects cooler 
than $T_{eff} \lesssim$1200~K \citep[defyning the L/T dwarf boundary;][] {Bur06}. 
This technique has been successfully applied to a search for T-dwarfs in the young IC~348 cluster \citep{Bur09}.

This paper is organized as follows. Section~\ref{obs} describes the WIRCam observations in the Serpens Core and the catalog extraction procedure. 
In Sects.~\ref{sel}-\ref{sect_par}  we use this catalog to identify possible T-dwarfs on the basis of methane absorption bands 
and estimate their intrinsic properties (T$_{eff}$, A$_V$, etc.). 
In Sect.~\ref{discussion} we investigate the nature of our T-dwarfs candidates using complementary near to mid-IR data and T-dwarf spectral models.
The conclusions of this work are given in Sect.~\ref{concl}.

\section{Observations and data reduction \label{obs}}

The imaging observations presented in this work were obtained  as part of a large CFHT (Canada-France-Hawaii Telescope) 
key program (08AF98) aimed at the characterisation of the low-mass population of several young star-forming regions (P.I. J. Bouvier). 
The data were obtained in queue-scheduled observing mode between 20 April and 19 May 2008 using the 
Wide-field InfraRed Camera (WIRCam) at CFHT. 
The WIRCam consists of four Hawaii-II2-RG 2048$\times$2048 array detectors 
with a pixel scale of 0.3$^{\prime\prime}$ or 0.15$^{\prime \prime }$ with microdithering \citep{Pug04}. 
The four detectors are arranged in a 2$\times$2 pattern, with a total 
field of view of 21.5$^\prime \times$21.5$^\prime$ and small gaps of 45$^{\prime\prime}$ between adjacent chips. 
The average CCD read-out noise and gain are 30~e- and 3.8~e-/ADU, respectively. 

In this paper, we focus on the observations of the Serpens core cluster 
obtained through the CH$_4$off ($\lambda_C$=1.58~$\mu$m, $\Delta \lambda$=0.1$\mu$m) and CH$_4$on ($\lambda_C$=1.69~$\mu$m, 
$\Delta \lambda$=0.1$\mu$m) filters, along with observations through the $JHK_S$ broad-band filters obtained using micro-dithering. 
A single WIRCam pointing centered at  RA=$18^h 29^m 57^s$ and Dec=$+01^d 13^m 08^s$ 
was sufficient to cover the Serpens core cluster. 
Each tile in each filter was observed using a 7-point dithering pattern 
selected to fill the gaps between detectors and accurately subtract the sky background. 
Exposure times, seeing and air mass conditions are summarized in Table~\ref{sum_obs}.

\begin{table}
\caption{Journal of the observations.}           
\label{sum_obs}      
\centering                       
\begin{tabular}{lllll}      
\hline\hline               
Field &  Date       & Filter & T$_{exp}$ & Seeing                 \\   
      &  (d/m/y)      &        & (sec)     & ($^{\prime\prime}$)          \\   
\hline                        
Serpens Core  &  17-19/05/2008  & CH$_4$off &   2520     &  0.6     \\  
 	                  &  17-19/05/2008  & CH$_4$on  &  7560      &  0.6     \\ 
 	                  &  20/04/2008       & $J$                   &    8405    &  0.7     \\
 	                 &   20/04/2008      & $H$                  &     1405   &  0.7     \\ 
 	                 &   20/04/2008      & $K_S$      &       1405  &  0.7     \\ 
\hline                                   
\end{tabular}
\end{table}

Individual images were primarly processed using the standard CFHT Data Processing and Calibration Pipeline for WIRCam (I'iwi; Albert et al., in preparation), 
which includes bias subtraction, flat-fielding, non-linearity correction, cross-talk removal, sky subtraction, and astrometric calibration. 
The astrometric World Coordinate System (WCS) solution is purely linear (i.e., assumes a constant scale) and is determined in two steps.
Knowing the geometry of the detectors (WIRCam has a field distortion of about 1\% in the corner of the mosaic), 
a first fit using all 2MASS stars found on the whole mosaic is performed. Then, provided enough stars are found, 
the same fit method operates on each individual detectors to refine the WCS solution individually. 
This generally yields an astrometric accuracy of $\sim$1 arc second  and the RMS scatter 
of the resulting WCS solution is generally about 0.5 arc second\footnote{For further details, we defer the reader to the WIRCam webpage: 
\emph{http://www.cfht.hawaii.edu/Instruments/Imaging/WIRCam/IiwiVersion1Doc.html\#Part5.}}.

Afterwards, the combination of individual exposures into final stacked images and their photometric calibration  were performed using the SCAMP and SWarp standard packages 
by Terapix, the data reduction centre at the Institut d'Astrophysique de Paris, France \citep[see][and references therein]{Mar07}.
The photometric calibration of the WIRCam data in the broad-band filters is done as part of the nominal pipeline reduction using 2MASS stars 
in the observed frames and images are renormalized to an arbitrary photometric zero-point of 30~mag.
Typical estimated errors in the WIRCam zero point determination are of the order of 0.05~mag, depending on 
the number of 2MASS stars available, on the variability of those stars and on small problems in flat-fielding and/or sky subtraction from night to night. 
A further check/refinement of the extracted $JHK_S$ magnitudes was performed using 2MASS photometry for the brightest stars in the field, 
as is described in the next section (i.e., Sect.~\ref{cat}). Images in the methane filters have no external photometric calibration and the CH$_4$off and CH$_4$on 
magnitudes are given here on an arbitrary albeit internally consistent scale, 
so that H-CH$_4$on=H-CH$_4$off=0 for unreddened field dwarfs.

\subsection{Photometric catalog \label{cat}}

The source extraction and photometry were performed by using a 
combination of SExtractor \citep{Ber96} and PSFex \citep{Ber11}. 
SExtractor extracts well defined stellar-like objects, which are used by PSFex to compute a PSF 
model that is allowed to vary with position on the detector. Then, SExtractor uses this PSF model 
to accurately extract and measure the photometry of all the sources detected on the image. 

\begin{figure*}
\centering
\includegraphics[width=13cm]{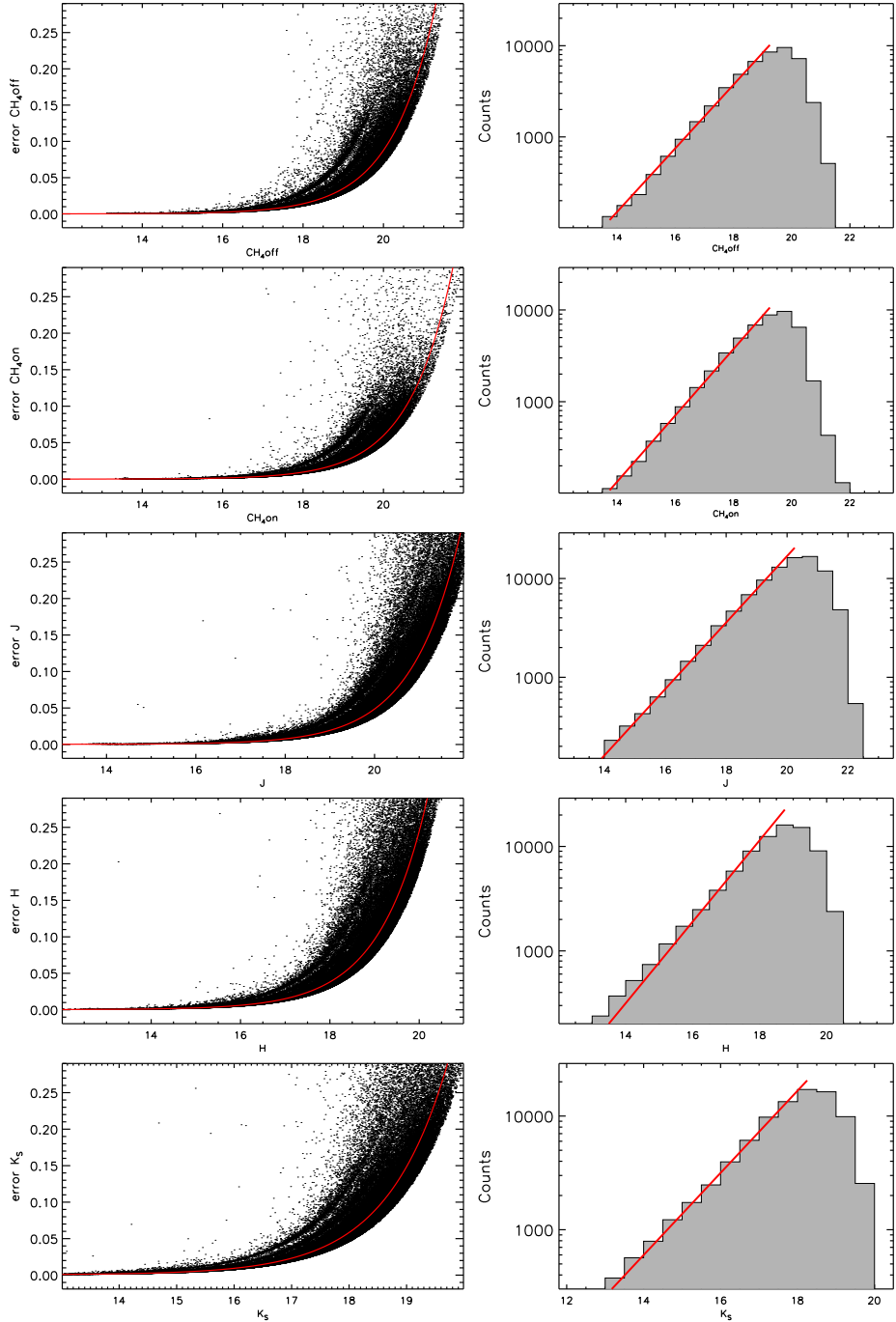}
\caption{{\bf Left panels:} Photometric errors as a function of PSF magnitudes and relative exponential fit (continuous line) 
for all the point-like sources detected in the CH$_4$off, CH$_4$on and $JHK_S$ images. 
{\bf Right panels:} Number of detection as a function of magnitude. The line shows the linear fit to the histogram points used 
to find the turning point of the distribution, indicating our completeness limit. }
\label{fig_errors}
\end{figure*}

T-dwarf candidates are expected to be fainter and, hence, possibly missed in the CH$_4$on image because of their CH$_4$ absorption. 
Thus, we first perform the source extraction on the CH$_4$off image and then use this catalog as 
input list when running Sextractor on the CH$_4$on image. 
To ensure the detection of all the faint sources present in the images 
the extraction criteria used are not too stringent. An object is extracted if it complies with the requirement 
to have three contiguous pixels (DETECT\_MINAREA=3) with fluxes 1.5 times above the background variation
(DETECT\_THRESH =ANALYSIS\_THRESH=1.5). In order to minimize the number of spurious detections, 
the SExtractor cleaning efficiency parameter was set to the maximum value (CLEAN\_PARAM=1). An inspection of the 
images and detections showed that all the objects seen by eye are detected. 
However, this visual inspection showed that a significant fraction of blended sources was missed. 
A higher selection threshold (DETECT\_THRESH =ANALYSIS\_THRESH=5) 
allowed us to recover those sources and we created a final master catalog combining the results of the two runs and 
containing $\sim$130~000 objects. 
The summary of saturation limits and limiting magnitudes, after cleaning for saturated objects and obvious spurious detections/artifacts 
(i.e. Sextractor FLAGS$\leq$2), is given in Table~\ref{tab_obs}. 
The approximate completeness limit in each filter was derived as the point where the histogram of the magnitudes not corrected for extinction (Figure~\ref{fig_errors})
diverge from the dotted line, which represents the linear fit to the logarithmic number of 
objects per magnitude bin, calculated over the intervals of better photometric accuracy \citep{San96,Wai92}.  

About 60\% of the objects detected in CH$_4$off and CH$_4$on images have a counterpart in the broad-band images. 
An analogous procedure was applied to obtain $JHK_S$ photometry. 
We perform the source extraction on the $J$, $H$ and $K_S$ images independently using Sextractor and adopting the same configuration parameters as for the 
methane filters, i.e. DETECT\_MINAREA=3, DETECT\_THRESH =ANALYSIS\_THRESH=1.5 and CLEAN\_PARAM=1. 
A second run was also performed on the $JHK_S$ images using an higher selection threshold (DETECT\_THRESH =ANALYSIS\_THRESH=5) 
to recover the blended sources.
The  final master catalog, obtained by combining the results of the two runs, 
contains $\sim$94000 sources for the $J$ filter, $\sim$81000 for the $H$ filter and $\sim$87000 for the $K_S$ filter. 
Approximately 4000 of the brightest stars in our field have a counterpart in the 2MASS catalogue. 
We used this sample to check the photometric accuracy and pipeline reduction of the WIRCam images and to 
estimate the difference between the CFHT and 2MASS photometric systems. 
The mean magnitude differences between the two systems is found to be 0.07, 0.06, and 0.1 mag for J, H, and $K_S$, respectively, 
which is of the order of expected zero point uncertainty ($\sim$0.05, Sect.~\ref{obs}). 
The dispersion of the differences is $\sim$0.08~mag form the three filters and 
reflects the possible sources of error mentioned above (i.e., number  and variability of the available 2MASS stars, 
non-optimal flat-fielding and/or sky subtraction from night to night, etc.; Sect.~\ref{obs}). Given that the dispersion is as large as the typical 
magnitude difference between the two systems, we choose to not correct our photometry for these offsets. 
Moreover, this comparison hold for the brightest stars in our images, because fainter ones are undetected in the 2MASS survey.
This implies that color effects arising from CFHT filters can not be taken into account in this calibration, although they are expected to be significant 
because the WIRCam and 2MASS filters design differ substantially.
Thus, throughout this work, all the WIRCam J, H, and K$_{S}$ photometry is given in the CFHT Vega system. 
The saturation limits and limiting magnitudes in the broad-band filters, after cleaning for saturated objects and obvious spurious detections/artifacts 
(i.e., Sextractor FLAGS$\leq$2), are given in Table~\ref{tab_obs}. 
The approximate completeness limit in each broad-band filter was derived as for the methane filters (Figure~\ref{fig_errors}). 

The catalogues from all filters are combined into a single database by requiring a positional match better than 1$^{\prime\prime}$, 
corresponding to the astrometric accuracy of our images (see Sect.~\ref{obs}).

\begin{table}
\caption{Saturation limit, limiting magnitudes at the 10$\sigma$ and 5$\sigma$ level and completeness limit  in each filter. }           
\label{tab_obs}      
\centering                       
\begin{tabular}{lllll}      
\hline\hline               
Filter & Saturation & Mag$_{10\sigma}$ & Mag$_{5\sigma}$ & Completeness  \\   
       &  limit      &                 &                 &  limit \\   
\hline                        
CH$_4$off &  13.0 & 20.1 & 20.9 & 19.5 \\   
CH$_4$on  &  13.5 & 20.6 & 21.3 & 19.5 \\ 
$J$                   &  13.5 & 20.8 & 21.5 & 20.5 \\   
$H$                 &  12.5 & 19.0 & 19.8 & 19.0 \\   
$K_S$        &  12.5 & 18.6 & 19.3 & 18.5 \\   
\hline                                   
\end{tabular}
\end{table}

\section{T-dwarf candidate selection \label{sel}}

As demonstrated by \citet{Bur09}, the flux ratio between the CH$_4$on and CH$4$off filters gives a good measure 
of the strength of the methane absorption in T-dwarfs. 
Specifically, the CH$_4$on filter probes the CH$_4$ absorption band centered at $\sim$1.66~$\mu$m and typical of T-dwarfs, 
while the  CH$_4$off filter covers a wavelength range relatively featureless in T-type objects and, hence, can be used as the primary continuum filter.

\begin{figure}
\centering
\includegraphics[width=9cm]{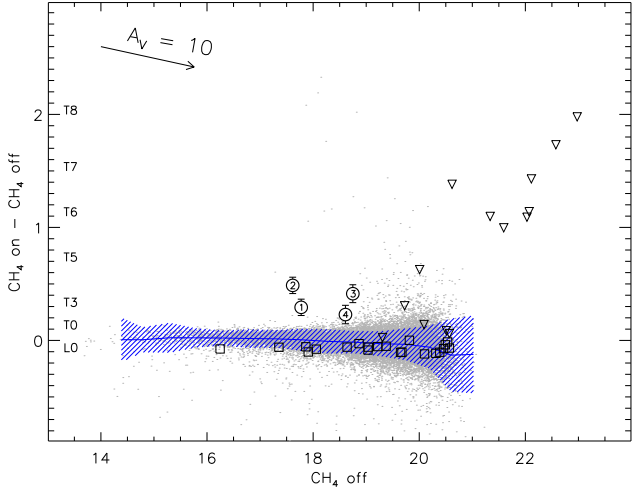}
\caption{CH$_4$on-CH$_4$off vs. CH$_4$off diagram for the point-like objects detected in Serpens core (small grey dots). 
The continuous line and the line-filled area show the median locus of field stars and their dispersion, respectively. 
The squares and triangles show the location of confirmed field L and T dwarfs, respectively; 
their magnitudes have been shifted to the distance modulus of Serpens. 
The circles show the position of our the T-dwarf candidates; their ID number (see Table~\ref{tab_phot}-\ref{tab_par}) is indicated and 
errors bars are within the circle size. The extinction vector is shown for A$_V$=10~mag. 
The CH$_4$on-CH$_4$off vs. spectral type calibration relation by \citet{Bur09} is also reported on the y-axis.}
\label{fig_sel}
\end{figure}

As shown by \citet{Bur09} in their Figure~2, the CH$_4$on-CH$_4$off color in the CFHT methane filters 
is roughly equal to zero for L-type and earlier type dwarfs and smoothly increases towards later spectral type, 
so that T-dwarfs have colors above $\sim$0.1.
Figure~\ref{fig_sel} shows the CH$_4$on-CH$_4$off against CH$_4$off diagram for point-like objects detected in the observed area in Serpens core.
We use this diagram to select T-dwarf candidates following the steps described below:

\begin{itemize}

\item Most of the stars in our field have spectral type earlier than L, as demonstrate by the over-density of points at 
CH$_4$on-CH$_4$off$\approx$0 in Figure~\ref{fig_sel}. We use the median CH$_4$on-CH$_4$off color of these stars as a function of the CH$_4$off magnitude 
to define the reference template with respect to which the CH$_4$ absorption feature is sought. 
The continuous line and the line-filled area in Figure~\ref{fig_sel} show the median locus of field stars and their dispersion ($\sigma$), respectively.

\item We consider the photometric uncertainty on the CH$_4$on-CH$_4$off color ($\delta = \sqrt{\Delta CH_4 on^2 + \Delta CH_4 off^2}$) and 
select a first sample of T-dwarf candidates by considering all those objects with (CH$_4$on-CH$_4$off)-$\delta$ color 
exceeding that of the median locus of field stars by more than 3$\sigma$. The objects just below this limit could still be early T-dwarfs but it would be more 
difficult to extract them because of photometric errors. With this criterion $\sim$200 sources were selected. 

\item This sample was visually inspected on both the CH$_4$off and CH$_4$on images and the $JHK_S$ images. 
Most of the objects turned out to be nebulous detections, ghosts or detector cross-talk, spurious detections/artifacts in the vicinity of saturated objects or image edges. 
Thus, the initial sample was reduced to a shortlist of only 4 point-like candidates. 

\end{itemize}

\begin{figure*}
\centering
\includegraphics[width=12cm,angle=-90]{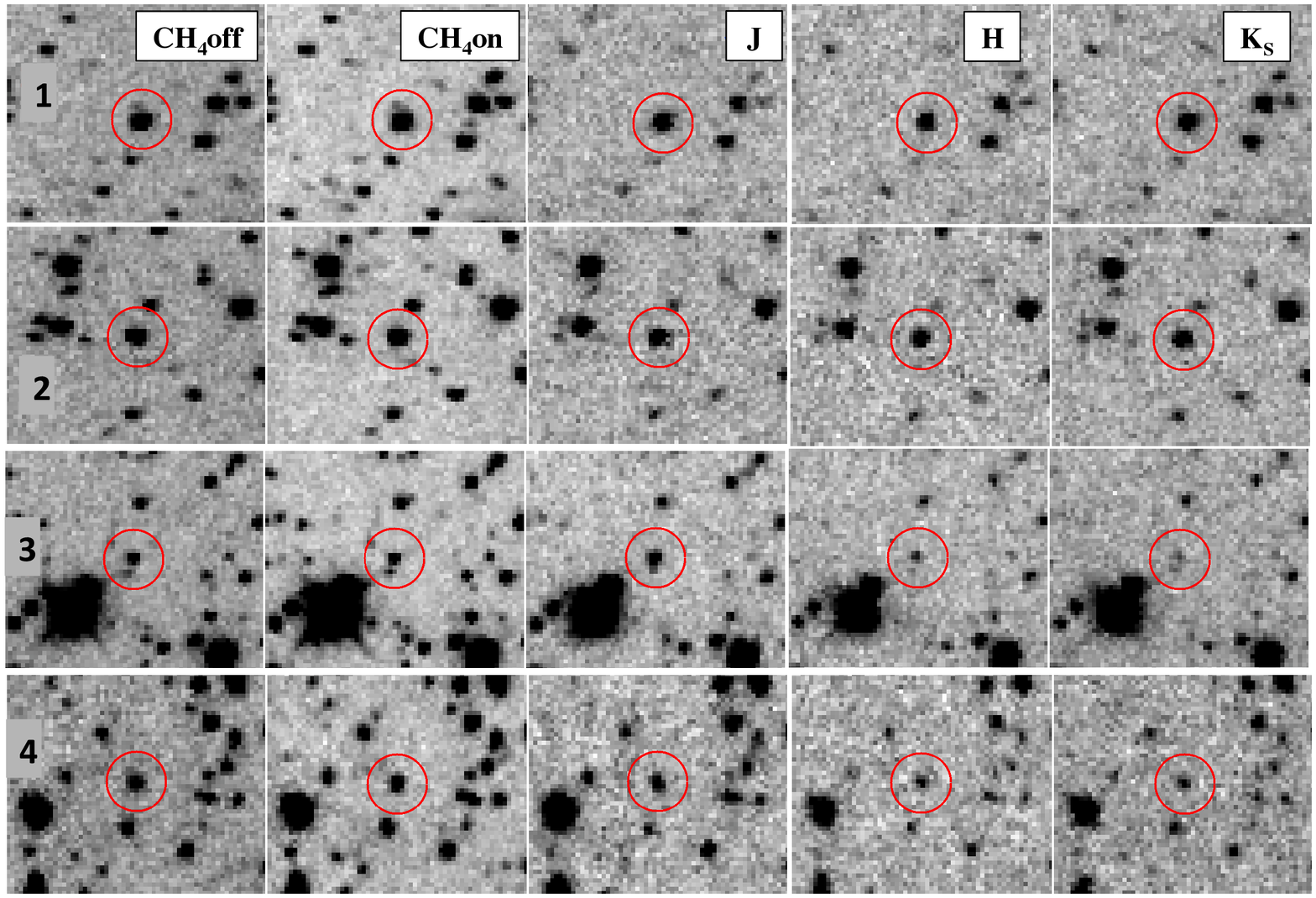}
\caption{Images of the four T-dwarfs candidates in each of the WIRCam@CFHT filter used in this work, as indicated in the labels. 
Each snapshot covers an area of 18$^{\prime\prime} \times$ 15$^{\prime\prime}$; north is up and east to the left.}
\label{im_TDcand}
\end{figure*}

The PSF photometry of these four T-dwarf candidates in the CH$_4$on/CH$_4$off filters and in $JHK_S$ is listed in Table~\ref{tab_phot}.
Thumb-sized images of the four candidates are shown in Figure~\ref{im_TDcand}.

\begin{table*}
\caption{Photometry of the T-dwarf candidates. CH$_4$on and CH$_4$off magnitudes are in the WIRCam 
system, while $riJHK_S$ photometry is given in the Vega system.}           
\label{tab_phot}      
\centering                       
\begin{tabular}{llllllllll}      
\hline\hline               
ID   &   R.A. J2000    & Dec. J2000  & $r_{CFHT}$ & $i_{Subaru}$  &  CH$_4$off  & CH$_4$on &  $J$ & $H$ & $K_S$  \\ 
     &  (hh:mm:ss)     & (dd:mm:ss)   &                        &         &                        &                     &          &          &      \\ 
\hline                        
1 & 18:29:56.58  & +01:18:35.93   & --                            & 22.98$\pm$0.03  &   17.77$\pm$0.05  & 18.07$\pm$0.05  & 19.10$\pm$0.05  &   17.71$\pm$0.06  &  16.99$\pm$0.05  \\ 
2 & 18:29:57.64  & +01:19:40.61   & --                            & 22.67$\pm$0.03  &   17.61$\pm$0.05  & 18.10$\pm$0.05  & 18.84$\pm$0.05  &   17.64$\pm$0.06  &  17.07$\pm$0.06  \\  
3 & 18:30:27.89  & +01:14:52.22   & --                            & 23.98$\pm$0.10  &   18.75$\pm$0.05  & 19.16$\pm$0.06  & 19.38$\pm$0.06  &   19.08$\pm$0.16  &  18.90$\pm$0.27  \\  
4 & 18:30:37.24  & +01:18:37.68   & 23.57$\pm$0.07 & 22.14$\pm$0.02  &   18.61$\pm$0.06  & 18.84$\pm$0.06  & 19.71$\pm$0.07  &   18.67$\pm$0.09  &  18.25$\pm$0.10  \\ 
\hline                                   
\end{tabular}
\end{table*}

\section{Reddening and spectral type estimates \label{sect_par}}

Knowledge of the candidates' extinction (A$_V$) is required in order to be 
able to estimate their spectral type and collect further information on their nature.

\begin{figure*}
\centering
\includegraphics[width=18cm]{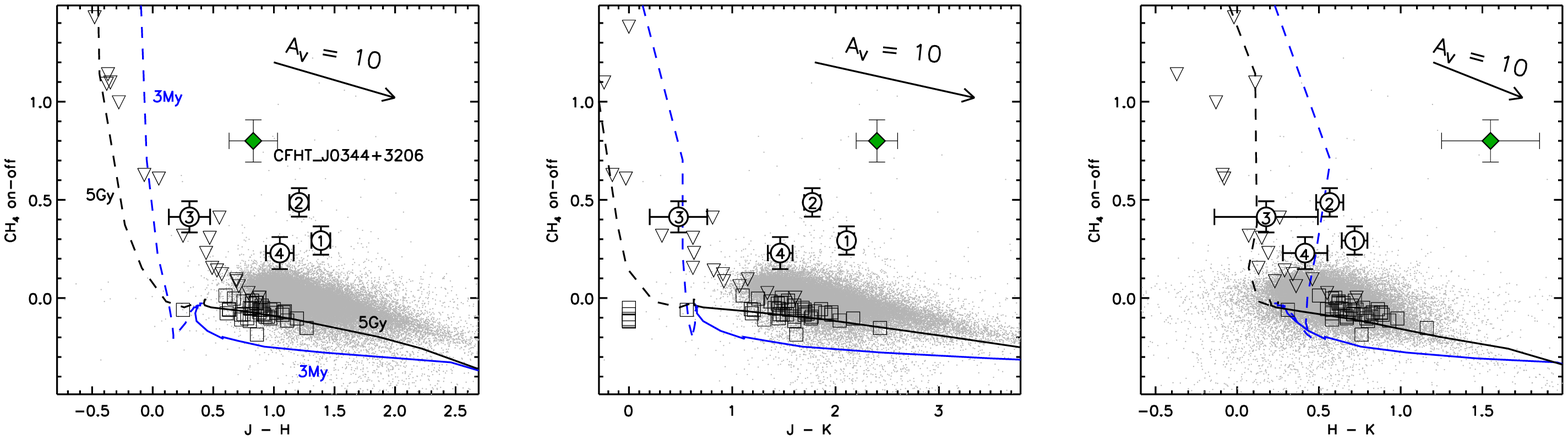}
\caption{CH$_4$on-CH$_4$off vs. $J-H$, $J-K$ and $H-K$ for the point-like objects detected in Serpens core (small grey dots). 
The circles with error bars are the T-dwarf candidates in this study; their ID number is indicated. 
The squares and triangles show the location of confirmed field L and T dwarfs, respectively \citep{Chi06, Gol04,Kna04}. 
The position of the CFHT\_J0344+3206 mid-T dwarf candidate (green diamond) is plotted for comparison. 
The lines show the 5~Gyr and 3~Myr COND isochrones (dashed lines) and the 5~Gyr and 3~My DUSTY isochrones (continuous line), 
as indicated in the labels. The extinction vector is shown for A$_V$=10~mag.}
\label{CMD_ext}
\end{figure*}

We do not know \emph{a priori} whether our candidates are field T-dwarfs or young members of Serpens. 
In both cases, we cannot give accurate A$_V$ value for each object based only on photometry, because we do not know their distance; 
they could be foreground objects closer than Serpens, cloud members located between 260 and 415~pc, i.e. the distance range 
of Serpens \citep[see Table~2 by][]{Dzi10} or even being located beyond the cloud. 
Thus, we estimate the range of possible extinction values for each candidate as follows:

\begin{itemize}

\item According to the study by \citet{Cha09}, the mid-infrared extinction law in Serpens is consistent with the \citet{Wei01} R$_V$= 3.1 
diffuse interstellar-medium dust model below A$_K \lesssim$1 (A$_V \lesssim$9). We use this prescription to define the direction of the extinction vector.

\item Assuming that our candidates belong to Serpens, extinction can be estimated using color-color diagrams of 
CH$_4$on-CH$_4$off versus $J-H$, $J-K_S$ and $H-K_S$, as plotted in Figure~\ref{CMD_ext}. 
The extinction is computed for each candidate using the extinction vector and regressing the objects back towards 
the 3~Myr PMS isochrone \citep{Cha00}, i.e. the typical age of the Serpens cloud core population \citep{Win09,Oli09}. 
The extinction values derived through this method (A$^{3My} _V$) are reported in Table~\ref{tab_par}. 
 
\item R. Gutermuth (private communication) calculated an extinction map of the Serpens core using 2MASS data and the method explained in \citet{Gut07}. 
This map has a spatial resolution of 25$^{\prime\prime} \times$25$^{\prime\prime}$ and entirely covers the area observed by us. 
By comparing this map with our WIRCam images, we estimate that the crowding of our field is such that typically $\sim$50 stars fall in the map resolution box. 
This number is suitable for a statistically significant estimate of A$_V$ in the given box/direction based on the \citet{Gut07} method.
Because the extinction map takes into account all the extinction occurring along the line of sight, 
it can be used to estimate the maximum extinction towards our candidates under the assumption that they are located beyond the Serpens cloud. 
Moreover, the extinction assigned to each object (A$^{map} _V$ in Table~\ref{tab_par}) is the average value in a 
25$^{\prime\prime}$ box around the object itself and, as such, can be inaccurate 
for very young and highly embedded objects, which present peculiar extinction due to the material in their envelope/disk. 
In Sect.~\ref{Spitzer} we will see that this is not likely to be the case for our candidates. 
However, the average A$_V$ value computed in a given 25$^{\prime\prime}$ box can be still inaccurate for an object located in a specific position within the box, 
because of the effect of small scale differential reddening, presence of cloud filaments, etc.

\item Finally, to estimate the lower limit to the extinction of our candidates, we assumed that they are foreground T-dwarfs located within 260~pc of our Sun, 
i.e. the minimum distance estimated for Serpens. The typical diffuse absorption is taken to be 0.7-1~mag/kpc 
in the Solar Neighborhood \citep[see, e.g.,][]{Ich82}. Thus, interloping field objects between our Sun 
and Serpens would be subject to a 0$\lesssim$A$_V \lesssim$0.2~mag extinction. 

\end{itemize}

For each candidate, we considered a minimum extinction of 0$\lesssim$A$_V \lesssim$0.2~mag, the extinction occurring along the line of sight, given by the extinction map, 
and the extinction expected if the object belongs to the young population of Serpens (A$^{map} _V$ and A$^{3My} _V$ in Table~\ref{tab_par}, respectively) 
and calculated the relative dereddened CH$_4$on-CH$_4$off colors. 
We then used the CH$_4$on-CH$_4$off vs. spectral type calibration by \citet{Bur09} (see their Figure~2) to estimate the spectral type range 
consistent with these dereddened colors. 
Note that the calibration relation by \citet{Bur09} is based on field dwarfs; the actual colors of our LT dwarfs might 
be slightly different because of the effects that reduced gravity has on the opacities \citep[see, e.g.,][and references therein]{Cas11}.

The spectral type estimates for our candidates are reported in Table~\ref{tab_par}, 
together with the corresponding effective temperature range according to the calibration relation by \citet{Vrb04}.

\begin{table}
\caption{Estimates of visual extinction, spectral type and effective temperature for our T-dwarf candidates 
based on their dereddened CH$_4$on$-$CH$_4$off colors. A$^{3My} _V$ is the visual extinction when 
assuming that the object belongs to the 3~My old population of Serpens, while A$^{map} _V$ is 
derived from the extinction map by R. Gutermuth (private communication).}           
\label{tab_par}      
\centering                       
\begin{tabular}{lllll}      
\hline\hline               
ID   & A$^{map} _V$  &   A$^{3My} _V$  & Spec. Type &  T$_{eff}$ \\ 
     &  (mag)        &      (mag)      &            &            (K)        \\ 
\hline                        
1  &   9.2   &    9.4    &	T2-T4 &  1390-1290   \\ 
2  &   7.9   &    7.1    &	T4-T5 &  1290-1190   \\ 
3  &   4.2   &    1.0    &	T3-T4 &  1360-1290  \\ 
4  &   2.8   &    5.5    &	T1-T3 &  1390-1360   \\ 
\hline                           	  
\end{tabular}
\end{table}

\section{On the nature of the T-dwarf candidates: are they young members of Serpens? \label{discussion}}
 
In this section we use color-magnitude diagrams (CMDs), color-color (CC) diagrams and spectral energy distributions (SEDs) 
to assess whether the optical/IR properties of our T-dwarf candidates are consistent with those 
of field objects or indicate the presence of IR excess emission typical of very young objects, 
which would in turn indicate membership to the Serpens core cluster. 
This analysis is performed using our WIRCam data in combination with $r$ and $i$-band 
imaging data from MegaCam at CFHT and Suprime-Cam at the Subaru telescope, and {\it Spitzer} imaging. 

\subsection{WIRCam near-IR color-magnitude and color-color diagrams \label{cmd_IR}}

\begin{figure*}
\centering
\includegraphics[width=18cm]{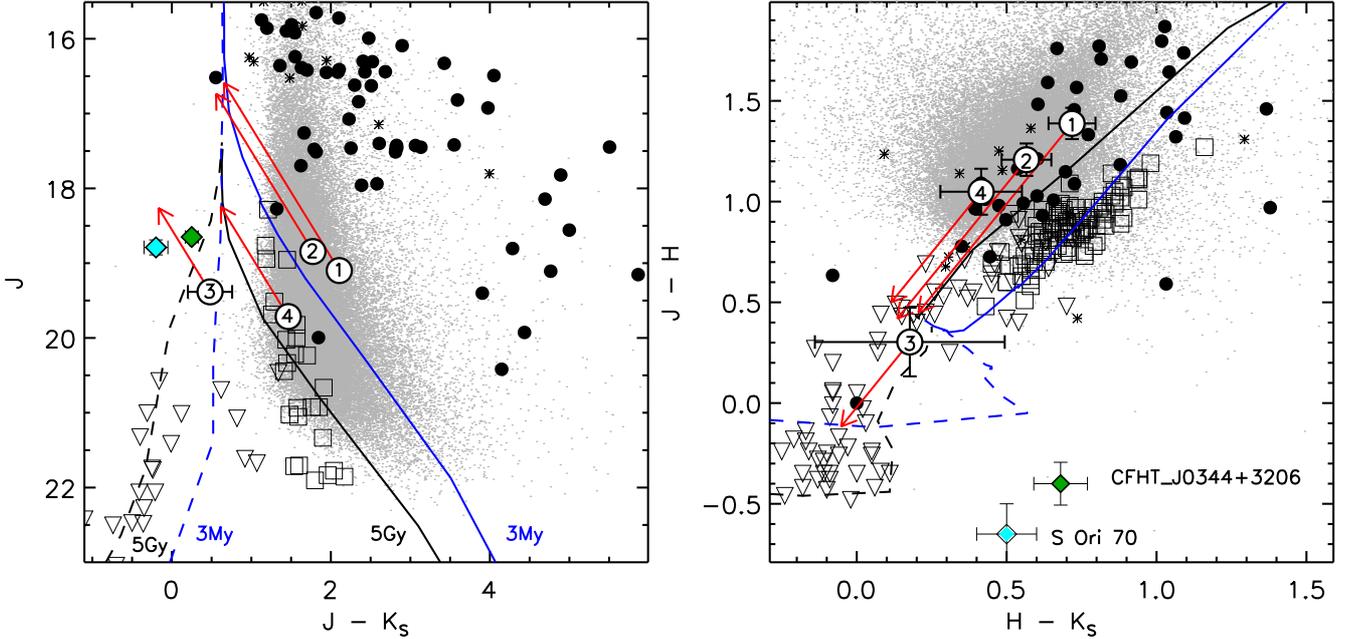}
\caption{$J$ vs. $J-K_S$ (left panel) and $J-H$ vs. $H-K_S$ diagrams (right panel) for the point-like objects detected in Serpens core (small grey dots). 
The circles with error bars are the T-dwarf candidates in this study; their ID number is indicated and the arrows indicate their position after correction for interstellar reddening.
The Serpens core class II (black dots) and class III (asterisks) YSO population,  the S~Ori~70 mid-T dwarf (sky-blue diamond) and the CFHT\_J0344+3206 
mid-T dwarf candidate (green diamond) are plotted for comparison. The squares and triangles show the location of confirmed field L and T dwarfs 
 \citep{Chi06, Gol04,Kna04}, respectively, shifted to the distance of Serpens. 
Lines show the 5~Gyr and 3~My COND isochrones (dashed lines) and the 5~Gyr and 3~My DUSTY isochrones (continuous line).}
\label{CMD_jhk}
\end{figure*}

Figure~\ref{CMD_jhk} shows the position of our candidates on the $J$ vs. $J-K_S$ CMD and the $J-H$ vs. $H-K_S$ CC diagram before 
and after correction for interstellar reddening. In order to draw the de-reddening vector for our candidates, we adopt the maximum extinction expected 
towards our candidates (i.e., the maximum between the A$^{map} _V$ and A$^{3My} _V$ value from Table~\ref{tab_par}); 
in this way, we display the maximum possible shift of their position in the CMD and CC diagram due to reddening effects.

We also plotted the 5 Gyr DUSTY and COND models and the confirmed field L and T dwarfs  \citep{Chi06, Gol04,Kna04} to highlight 
the position expected for the field dwarf sequence. The locus expected for the young members of Serpens is indicated by 
the 3~Myr DUSTY and COND models and the young stellar object (YSO) population confirmed by the {\it Spitzer} \emph{core to disk} (c2d) survey \citep{Win07,Har07}. 
Finally, we plot for comparison the faint mid-T type object S~Ori~70 found by \citet{Zap02} 
towards the direction of the young $\sigma$~Orionis cluster and the dereddened position of CFHT\_J0344+3206, a 
mid-T dwarf candidate found by \citet{Bur09} in the direction of IC~348. 
The DUSTY and COND models and the positions of the field LT-dwarfs, S~Ori~70 and CFHT\_J0344+3206 
are shifted to the minimum distance estimated for Serpens (i.e. 260~pc). 
Note that it is still unclear whether S~Ori~70 is a field brown dwarf or a young planetary mass member of $\sigma$~Orionis; 
however, proper motion and the near- and mid-IR colors measurements by \citet{Zap08} 
support its membership to the cluster, with an estimated mass in the interval 2-7~M$_\odot$. 

The position in both diagrams of two of our candidates, namely ID~1 and 2, is consistent with them being mid-T dwarfs.
However, their dereddened magnitudes are brighter than expected for young T-dwarfs at the distance of Serpens 
and, hence, they might be field objects. They suffer from significant interstellar extinction (Table~\ref{tab_par}) and therefore can not be foreground objects. 
As discussed by \citet{Del08}, the only background contaminants may eventually be high redshift quasi-stellar objects (QSO; z$\gtrsim$6) or heavily reddened low redshift QSOs, 
which appear point-like and share the same very red colors as cool dwarfs.
Moreover, high redshift/heavily veiled  QSOs can present a moderate methane-like absorption signal in the CH$_4$on-CH$_4$off difference due to the combinations of specific redshifts and emission lines \citep{Ric03}. 
In Sect.~\ref{Spitzer} we use additional imaging in the $i$ and {\it Spitzer} pass-bands to assess whether this is the case for  ID~1 and 2.

ID~3 does not present a significant interstellar extinction and  its near-IR colors are consistent with it being a field mid-T dwarf. 
Its position on the $J$ vs. $J-K_S$ diagram and spectral type are very similar to those of the mid-T type dwarf S~Ori~70 in the $\sigma$~Orionis cluster and the T-dwarf candidate CFHT\_J0344+3206 in the direction of IC~348.
However, it is not redder than the sequence defined by field T-type dwarfs in the $J-H$ vs. $H-K_S$ CC diagram, as young objects should be. 
Indeed, the K-band flux is sensitive to gravity and it is fainter for young objects, especially for late T-dwarfs, because of the strongly reduced gravity \citep{Zap08}. 
Since it is brighter than field objects at the distance of Serpens (Figure~\ref{CMD_jhk}, left panel), ID~3 may be a foreground field T-dwarf. 
  
The interpretation of the CMD and CC diagram for candidate ID~4 is controversial, because it lies in both diagrams 
in a ``transition region'' mainly populated by field dwarfs but where faint and/or highly veiled YSOs might still be found. 
It cannot be a foreground object as it is significantly extincted  (Table~\ref{tab_par}) and has near-IR dereddened colors consistent with it being an early T-dwarf. 
It is brighter than field objects (Figure~\ref{CMD_jhk}, left panel), but this is still consistent with it being young (the larger radius of YSOs make them brighter than older objects with the same spectral type). 
Thus, ID~4 may be a cluster member.

It is  worth noting that these very same conclusions can be drawn from the 
inspection of the CC diagrams involving colors in the CH$_4$ filters, which we presented in Figure~\ref{CMD_ext}.

\subsection{Complementary MegaCam, Suprime-Cam and {\it Spitzer} data \label{Spitzer}}

The Serpens core cluster was observed on 2010-06-11 in the $r$ and $i$ bands with {\it MegaCam} \citep{Bou03} at CFHT and 
on 2008-06-05 in the $i$-band with {\it Suprime-Cam} \citep{Miy02}, the wide field camera located at the prime focus of the {\it Subaru} Telescope.
For details about these observations and the relative data reduction and calibration, we defer the reader to Bouy at al. (2012, in preparation). 
In Table~\ref{tab_phot} we report the $ri$ photometry of our four T-dwarf candidates calibrated to the Vega system. 
Only one of our candidates (ID~4) was detected in both filters, while the remaining three objects (ID~1, 2 and 3) were detected only in the $i$-band.

We also searched for possible mid- to far-IR flux measurements of our T-dwarf candidates in the AKARI \citep{Yam11},  WISE \citep{Wri10} and {\it Spitzer} databases. 
The AKARI and WISE all-sky surveys turn out to be too shallow for the detection of our T-dwarf candidates. 
However, three of them (ID~1, ID~2 and ID~3) have been detected in the deeper {\it Spitzer} c2d 
Survey, which mapped Serpens together with other five nearby molecular clouds, 
using both IRAC and MIPS, the two imaging cameras on {\it Spitzer} \citep{Eva09}. 
ID~4, the faintest of our candidates, was not detected by the {\it Spitzer} c2d Survey. 
The IRAC and MIPS~24$\mu$m fluxes for ID~1, ID~2 and ID~3 are reported in Table~\ref{tab_c2d}. 
We also computed for these three candidates the SED slope between the 
K-band (2.2$\mu$m) and the MIPS band at 24$\mu$m ($\alpha_{[K \& 24\mu m]}$, Table~\ref{tab_c2d}), 
which is traditionally used to assess the IR class of YSOs \citep{Lad84}. The  $\alpha_{[K \& 24\mu m]}$ we obtain are upper limits to the actual SED slopes, 
because only flux upper limits are available at 24$\mu$m for our objects.

\begin{figure*}
\centering
\includegraphics[width=17cm]{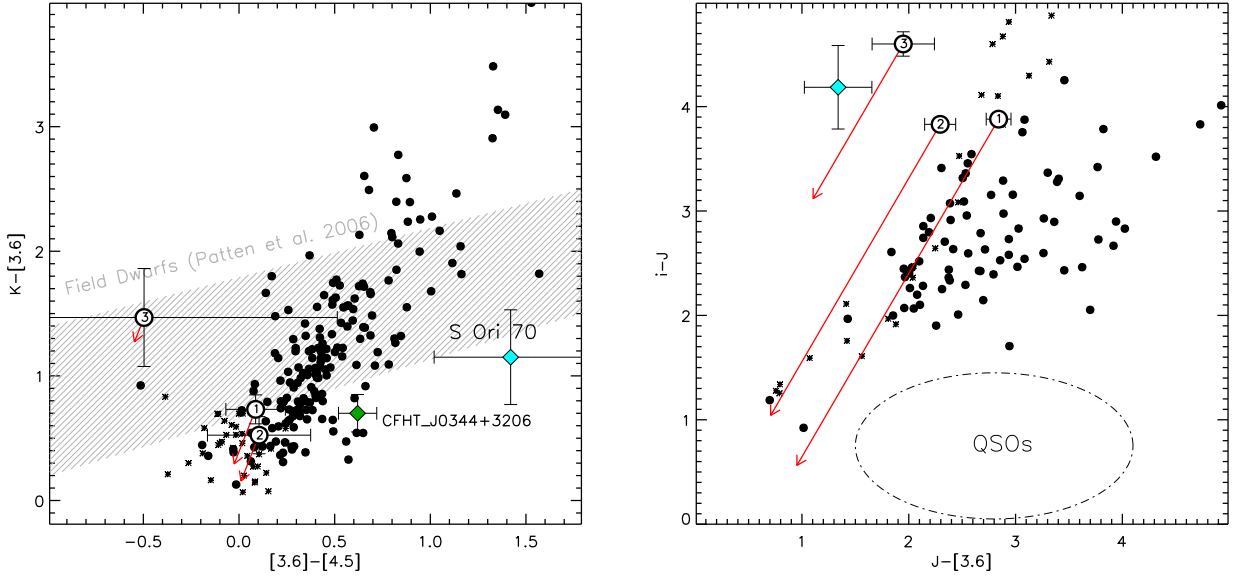}
\caption{{\bf Left panel:} $K_S - [3.6]$ vs. [3.6]-[4.5] color-color diagram. 
The circles with error bars are our the T-dwarf candidates; their ID number (see Table~\ref{tab_phot}-\ref{tab_par}) is indicated and the arrows indicate their position after correction for interstellar reddening. 
The Serpens core class II (black dots) and class III (asterisks) YSO population, the S~Ori~70 mid-T dwarf (sky-blue diamond), 
the CFHT\_J0344+3206 mid-T dwarf candidate (green diamond) and 
the field dwarf sequence from \citet{Pat06} (line-filled area) are plotted for comparison. 
{\bf Righ panel:}  $i-J$ vs. $J$-[3.6] color-color diagram. Symbols are as in the left panel. The dot-dashed ellipse marks the area occupied by QSOs in the SWIRE catalog \citep{Hat08}.}
\label{CMD_Spitz}
\end{figure*}

In Figure~\ref{CMD_Spitz} (left panel) we compare the position of our candidates in the $K_S - [3.6]$ vs. [3.6]-[4.5] CC diagram 
with the locus expected for field dwarfs \citep{Pat06}, the sequence defined by YSOs belonging to Serpens \citep{Har07,Win07} 
and the position of S~Ori~70 and CFHT\_J0344+3206. 

In Figure~\ref{CMD_Spitz} (right panel) we use the $i - J$ vs. $J$-[3.6] CC diagram to compare the colors of our candidates with the locus of
YSOs in Serpens, the position of S~Ori~70 and the colors expected for background QSOs, which might contaminate our sample. 
As demonstrated by \citet{Bou09}, very low mass YSOs (with or without mid-IR excess associated to the presence of a circumstellar disc) 
and QSOs occupy two very distinct areas of this diagram, thus it can be used to effectively separated the two populations. 

\begin{figure*}
\centering
\includegraphics[width=16cm]{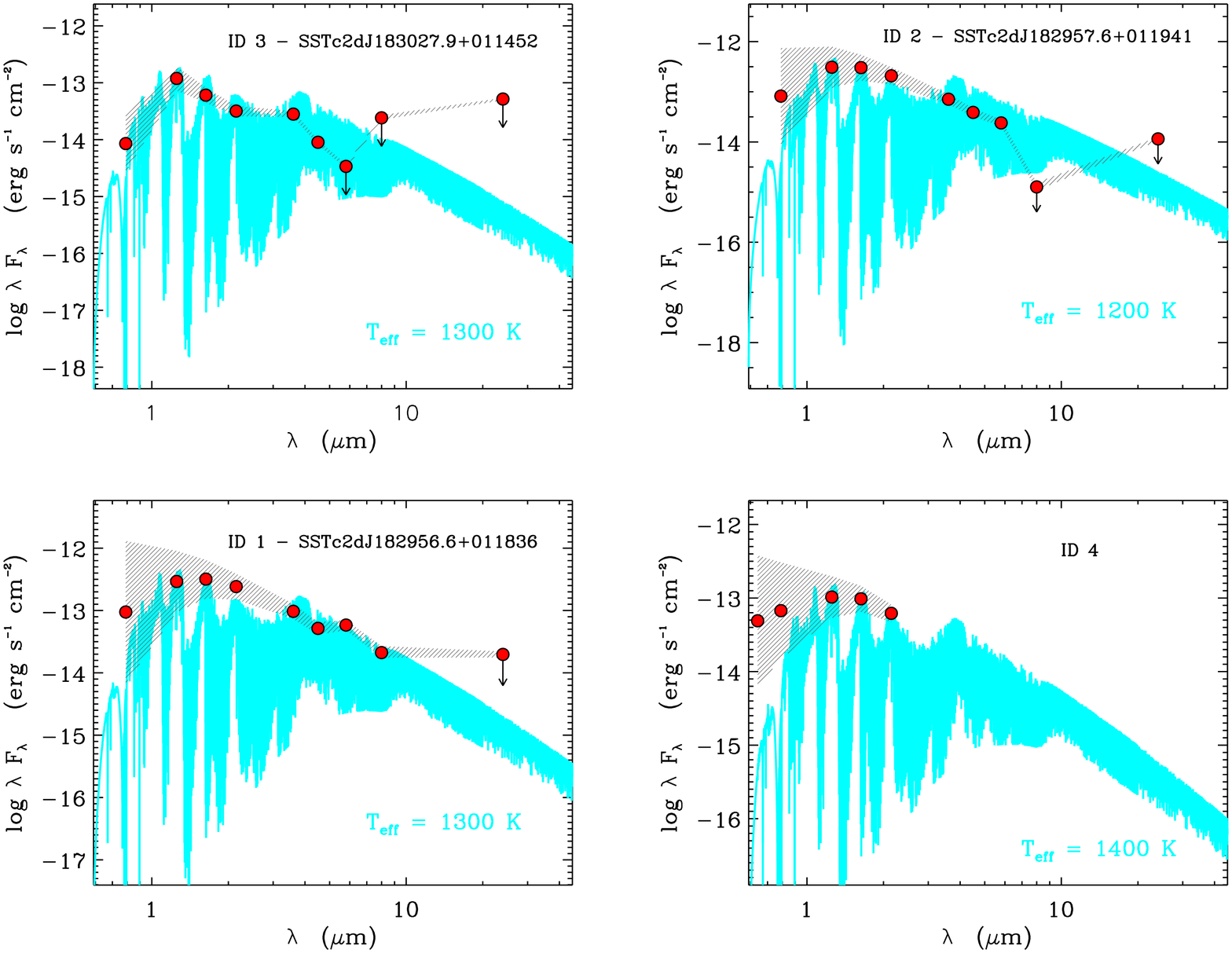}
\caption{Spectral energy distributions of the T-dwarf candidates in this study. Flux errors are smaller than the symbol size  and flux upper limits are marked with an arrow.
The line-filled area represents the flux variation depending on the adopted reddening correction. 
The oveplotted spectrum is the COND model by \citet{All01} and \citet{Bar03} with the same T$_{eff}$ as estimated for each object (indicated in the label and in Table~\ref{tab_par}) normalized to the $J$-band flux.}
\label{fig_SEDs}
\end{figure*}

Finally, in Figure~\ref{fig_SEDs} we present the SEDs of our four candidates from the optical up to 24$\mu$m. 
The line-filled area in each SED plot represents the flux variation depending on the adopted 
reddening correction (see Sect.~\ref{sect_par} and Table~\ref{tab_par}). 
The spectrum overplotted to each SED is the COND model spectrum by \citet{All01} and \citet{Bar03} with the same effective temperature as the object, 
normalized to the J-band flux; this model represents the pure stellar photospheric emission expected for an object of the given T$_{eff}$.

The {\it Spitzer} colors of ID~1 and ID~2 (Figure~\ref{CMD_Spitz}) are in general agreement with those of class III objects with thin or no disks. 
This is also in agreement with the shape of their SEDs, which might present a small IR excess emission only at 24~$\mu$m (Figure~\ref{fig_SEDs}). 
For both objects we find $\alpha_{[K \& 24 \mu m]} \lesssim -$1, which is typical of class~III objects with no prominent IR excess \citep{Lad84,Gre94}. 
Thus, the overall {\it Spitzer} colors/SED for these two objects indicate that they do not possess a clear IR excess emission typical of very young objects.
Their position in the $i - J$ vs. $J$-[3.6]  diagram indicates that their colors, when corrected for the estimated interstellar extinction (Table~\ref{tab_par}), 
approach the locus expected for QSOs. Since they are too bright to be high-redshift QSOs,  the most likely possibility is that they are heavily reddened low redshift QSOs.
Indeed, we estimated that the H$\alpha$ broad emission line of QSOs could contaminate the magnitude in the methane filters at  redfshift z$\sim$1.4 \citep{Ric03}. 

The SED of ID~3 (Figure~\ref{fig_SEDs}) might present IR excess starting from 8$\mu$m with a slope ($\alpha_{[K \& 24 \mu m]}$=0.27$\pm$0.04) typical of a class II YSO with a thick disk. 
However, only flux upper limits are available for $\lambda \ge$5.8$\mu$m and, hence, this  $\alpha_{[K \& 24 \mu m]}$ value is not robust.
The interpretation of the $K_S - [3.6]$ vs. [3.6]-[4.5] CC diagram (Figure~\ref{CMD_Spitz}) is  very uncertain because of the big error bars. 
However, the position of ID~3 appears to be consistent with the locus of field T-dwarfs. 
In particular, its $[3.6]-[4.5]$ color is bluer than expected for young mid-T dwarfs such as S~Ori~70 and CFHT\_J0344+3206. 
However, assuming that a thick disk surrounds this object, its $[3.6]-[4.5]$ color, peculiar for a young objects, might still be explained in term of 
different disk inclination with respect to S~Ori~70 and CFHT\_J0344+3206. Indeed, as shown in the top panels of Figure~7 by \citet{Rob06},
disk inclination significantly alter the SED shape at mid-IR wavelengths (3-10~$\mu$m) for very low-mass objects ($\lesssim$0.2~M$_\odot$). 
Finally, its position in the $i - J$ vs. $J$-[3.6]  diagram is clearly consistent with the YSO locus and, in particular, it is very close to the location of  S~Ori~70.
More robust imaging data at mid-IR wavelengths and/or follow-up spectroscopy  is needed to draw a firm conclusion about this object.

ID~4 is not detected in any of the {\it Spitzer} pass-bands and, hence, we can not assess the presence of a circumstellar disk around this object. 
For the sake of completeness, the optical/near-IR SED of ID~4 is shown in Figure~\ref{fig_SEDs}. 
The observed $i-J$ color of ID~4 (i.e., 2.43) is bluer than expected for early/mid T-dwarfs \citep[see, e.g., Table~3 by][]{Haw02}, 
while its observed  and dereddened $r-i$ color (1.43 and 0.82, respectively, adopting the Serpens extinction map (A$^{map} _V$ in Table~\ref{tab_par}) 
are consistent with the $r-i$ colors expected for early/mid T-dwarfs \citep[see, e.g., Table~3 by][]{Haw02}, in agreement with its near-IR colors (Figure~\ref{CMD_jhk}). 
Thus, our conclusions about this object remain as in Sect.~\ref{cmd_IR}: ID~4 may be a cluster member, although 
our data are not sufficient to conclusively prove this and spectroscopic follow-up is required. 


\begin{table*}
\caption{{\it Spitzer} fluxes and SED slopes ($\alpha_{[K \& 24\mu m]}$) for the three T-dwarf candidates in this study detected by the {\it Spitzer} c2d survey.}           
\label{tab_c2d}      
\centering                       
\begin{tabular}{llllllll}      
\hline\hline               
ID   &   c2d Name    & $\alpha_{[K \& 24 \mu m]}$ &  IRAC~3.6$\mu$m &  IRAC~4.5$\mu$m &  IRAC~5.8$\mu$m &  IRAC~8$\mu$m  & MIPS~24$\mu$m \\ 
     &               &                         &   (mJy)         &  (mJy)           &     (mJy)      &    (mJy)       &    (mJy)       \\
\hline                        
1  &  SSTc2dJ182956.6+011836 & $-$0.88$\pm$0.02  &  0.0869$\pm$0.0084  &  0.0610$\pm$0.0065  &  0.0930$\pm$0.0307  &  0.0464$\pm$0.0405 & $<$ 0.1410                 \\
2  &  SSTc2dJ182957.6+011941 & $-$1.06$\pm$0.02  &  0.0667$\pm$0.0084  &  0.0476$\pm$0.0102  &  0.0393$\pm$0.0403  &  $<$0.0028  	          &  $<$0.0836                  \\
3  &  SSTc2dJ183027.9+011452 &       0.27$\pm$0.04  &  0.0295$\pm$0.0078  &  0.0121$\pm$0.0108  &  $<$0.0060                   &  $<$0.0590	          &  $<$ 0.3930                 \\
\hline                                   
\end{tabular}
\end{table*} 
 
\section{Summary and conclusions \label{concl}}

We presented a deep methane imaging survey for planetary-mass T-dwarfs in the Serpens Core cluster. Our survey covers a field of view of  
21.5$^\prime \times$21.5$^\prime$ and about 127000 sources were detected in both the CH$_4$on and CH$_4$off filters.

We identified four potential T-dwarfs from their 1.6$\mu$m methane absorption band and used complementary $riJHK_S$ broad-band imaging and 
mid-IR flux measurements from the {\it Spitzer} c2d Survey to investigate their stellar and disk properties. 

Two of our candidates (ID~1 and ID~2) suffer from significant interstellar extinction and have near-IR colors similar to mid-T dwarfs. 
However, they are too bright to be planetary-mass members of Serpens and do not present the typical IR excess emission expected for young objects surrounded by a disk.  
Our analysis indicates that their dereddened near-IR colors approach the values expected for QSOs. 
In particular, since they are too bright to be high-redshift QSOs, they could be heavily reddened low redshift QSOs. 

Our candidate ID~3 does not present a significant interstellar extinction and its near-IR colors are more consistent with it being a field mid-T dwarf. 
It is brighter than field objects at the distance of Serpens and, hence, it may be a foreground T-dwarf. 
It is detected by the {\it Spitzer} c2d survey but only flux upper limits are available for $\lambda \ge$5.8$\mu$m and, hence, 
we can not assess the presence of a possible disk around this object. 
However, it is worth to keep this possibility open, because the position on the $J$ vs. $J-K_S$ and $i-J$ vs. $J$-[3.6] diagrams and the estimated spectral type of ID~3 are very similar 
to those of other young T-dwarf candidates, i.e. S~Ori~70 and  CFHT\_J0344+3206 in the direction of IC~348. 
 
Finally, ID~4 is significantly extincted, which exclude the possibility of it being a foreground object, and has dereddened 
optical and near-IR colors mainly consistent with those of early/mid T-dwarfs. 
It is brighter than field objects and, hence, it might be very young, as YSOs have larger radii and are brighter than older field objects of the same spectral class. 
Thus, ID~4 is a promising young T-dwarf candidate in our sample.

Considering the magnitudes of our candidates ID~3 and 4 (Table~\ref{tab_phot}) and their T$_eff$ (Table~\ref{tab_par}) based on the 
dereddened CH$_4$on$-$CH$_4$off colors, we estimated their masses 
assuming that they belong to the Serpens core cluster (i.e., age$\sim$3~Myrs and distance in the range 260-415~pc) and 
using both the COND and DUSTY evolutionary models by \citet{Cha00}. 
The estimated mass is in the range 2-4 Jupiter masses for both ID~3 and 4. 
Therefore, if they truly belong to the Serpens Core cluster, they would be 
amongst the youngest, lowest mass objects detected in a star-forming region so far. 

Follow-up spectroscopy is required to confirm the spectral type and reddening  of our four T-dwarf candidates and, hence, draw firm conclusions about their nature. 
Additional deep imaging at mid-IR wavelengths is needed to clarify the presence of a possible disk around ID~3.
 
\begin{acknowledgements}

We acknowledge financial support from IPAG ( (ANR 2010 JCJC 0501-1) and the Research and Scientific Support Department at ESA-ESTEC. 
We are grateful to H. Bouy, N. Huelamo and P. Delorme for many discussions during the data analysis, P. Harvey for explanation about the {\it Spitzer} c2d catalog 
and the anonymous referee for his careful reading and useful comments/suggestions.
We thank the QSO team at CFHT for their efficient work at the telescope and the data pre-reduction as well as the Terapix group at IAP for the image reduction. 
This work is based in part on data products produced and image reduction processes conducted at TERAPIX. 
This research has made use of the NASA/IPAC Infrared Science Archive, which is operated by the Jet Propulsion Laboratory, California Institute of Technology, 
under contract with the National Aeronautics and Space Administration. This research has also made use of the SIMBAD database, operated at CDS, Strasbourg, France.
The research leading to these results has received funding from the European Community's Seventh Framework Programme (/FP7/2007-2013/) 
under grant agreement No 229517.

\end{acknowledgements}

\bibliographystyle{aa}

\end{document}